\documentclass[12pt]{article}
\usepackage{epsfig,amsfonts,amssymb,graphicx,cite}
\usepackage[dvips]{color}

\begin{document}
\newcommand{\comment}[1]{}
\definecolor{Blue}{named}{Blue}
\definecolor{Red}{named}{Red}
\definecolor{Green}{named}{ForestGreen}
\definecolor{Black}{named}{Black}
\definecolor{Olive}{named}{OliveGreen}
\definecolor{Royal}{named}{RoyalBlue}
\definecolor{Orange}{named}{YellowOrange}
\begin{center}
{\Large \bf Story of a journey\\
Rutherford to the Large Hadron Collider and onwards..}\\
\vspace{0.5cm}
{Rohini M. Godbole}\\
\vspace{0.5cm}
{Theoretical Physics, CERN, CH-1211, Geneva-23,Switzerland\\
Centre for High Energy Physics, Indian Institute of Science, Bangalore, 560012, India.}
\end{center}
\begin{abstract}
In this article, I set out arguments why the Large Hadron Collider (LHC)
: the machine and the experiments with it, are a watershed for particle 
physics. I give a historical perspective of the essential link between 
development of particle accelerators and that in our knowledge of the laws 
governing interactions among the fundamental particles, showing how this 
journey has reached destination LHC. I explain how the decisions for the 
LHC design; the energy and number of particles in the beam, were arrived at.
I will end by discussing the LHC physics agenda and the time line in which the 
particle physicists hope to achieve it.
\end{abstract}
\section{Introduction}
More than a year and half  ago the `Large Hadron Collider' (LHC), came into 
limelight, due to the spectacular show of its start up, the equally spectacular
accident soon after and also for the doomsday stories that circulated around 
its start up.
The accident  put it out of action for a while. Now the necessary repairs done, the damaged pieces replaced, the machine has taken the first tentative steps 
in its life beginning on November 23, 2009. Having set a world record for the 
proton beam energy, of 1180 GeV, on November 30, 2009, it went back to the  
lower beam energy of $450$ GeV per beam  and ran for about two weeks in the 
collision mode. The detectors collected data, corresponding
to a few hundred thousand $pp$ collision events,\footnote{At full throttle 
LHC detectors will have to deal with over 600 million proton collisions
per second.} over this period  
confirming  that their intricate machinery performed as it should and can
measure the properties of the myriad of particles produced in these collisions,
to the desired accuracy.  After a winter shut down (partially caused  by the 
escalation in energy costs due to the increased energy demands otherwise 
in winter) the  machine started working again on 20 th February 2010. 
After circulating proton beams at the higher energy of 3.5 TeV since 19th 
March 2010, finally  collisions at the total centre of mass energy of
7 TeV happened  on the 30th  March 2010;  albeit this was still only
half the originally planned  energy and with 10 times less number 
of particles per bunch than initially foreseen for the restart.  The decision
to do so being taken, in the words of CERN Director General, Rolf Heuer, as a 
`prudent step by step approach' as `the LHC is not a turnkey project'. Papers,
presenting results from the December runs have already been published in 
research journals~\cite{:2009dt,Khachatryan:2010xs,Aad:2010rd} and more have 
been submitted already from the collisions in March~\cite{Khachatryan:2010us},
with thousands of authors signing them. Particle physicists: experimentalists 
and theorists alike, are waiting now with baited breath, just as they have 
been for more than a decade now, for the results which will come out of these 
collisions. I write these lines  from CERN, where even in the cafeteria now
the television screens display information about the LHC machine operation!

While the rest of the world came to know of this extremely complicated
endeavor only about a year and half ago, for the worldwide community of 
particle
physicists (which incidentally developed the World Wide Web (WWW) two decades 
ago), this is perhaps the concluding chapter of the long running love story 
between the world of fundamental constituents of matter and that of the 
accelerators which get these particles to move at the speed of light. In this 
article I discuss why thousands of particle physicists the world over, 
together, have embarked upon this truly mammoth project and worked  on it over 
three decades. I wish to highlight what we all hope to learn from doing 
experiments with this unique machine. The construction of this accelerator and 
that of the equally huge and complex detectors,
by itself, has been an amazing and utterly impressive engineering achievement. 
To appreciate the scope of this achievement by accelerator physicists and 
engineers, from CERN and  the rest of the world, including India, 
we should follow a story which began perhaps a century ago. It is also a story
of a development of a methodology of the ''mega' science projects that the
high energy physicists have developed.

The development of our knowledge of the laws of physics that function at the 
heart of the matter has been very closely interlinked with that of the 
accelerators. It may be said that the discovery of the first fundamental
particle, the electron, by J.J. Thompson in 1897 was due to the first
`accelerator: a cathode ray tube', which accelerated electrons.  The partnership
between accelerators and particle physicists on this adventure has continued 
through the century. The legendary physicist Lord Rutherford,  who discovered   
existence of a nucleus inside an atom using the alpha particles emitted by 
the radioactive nuclei, had said,  dreaming about high energy particles to 
uncover nature's secret ``It has long been my ambition to have
available a copious supply of atoms and electrons which have energies
transcending those of the alpha, beta  particles from the radioactive bodies''. 
His dream was full filled by Walton and Cockroft in the Cavendish Laboratory.
The target of energy (MeV : Million electron volt) to which the particles 
needed to be accelerated, was set by Gamow's theory of radioactive decay. 
This gave the height of Coulomb barrier to $\alpha$ emission in a nucleus
and hence indicated the energy level to which particles need to be accelerated 
for possible artificial radioactivity:  the aim of Rutherford's experiments at 
that time. Since then the nuclear/particle physicists have been setting the 
bar higher and higher and accelerator physicists have been clearing it with 
regularity, like Sergey Bubka or Yelena Isinbayeva.

Compared to the early machine by Walton and Cockroft which fitted in a room 
or the early cyclotron by Lawrence and Livingstone which was only 11 
inches in diameter, today's machines are truly gigantic. Further, the energies 
to which these machines accelerate the particles are higher by many orders of 
magnitude.  The Tevatron at the Fermi  National Accelerator 
Laboratory (FNAL) in Batavia in USA, held the record for accelerating 
protons to about a thousand times its rest mass energy $\sim$ 1000 GeV 
or 1 TeV, corresponding to a total collision energy of 2000 GeV; a feat that 
the LHC eclipsed by first reaching a total collision energy of 2360 GeV and 
then 7000 GeV, as said before.

Acceleration of particles to these high energies is not achieved by the 
use of electrostatic force alone, but has to be by a judicious combination
of the alternating electric fields and magnetic fields as is done in  the
more sophisticated cyclotron. Lawrence's first model, costing only US \$ 25 
in 1931, used 2000 volts of electricity but produced protons with energy 
80,000 electron volts. Magnetic fields are not only required to facilitate 
this acceleration of particles, but later also to keep them on a tight leash 
and steer the beam  around its path as a tight bunch, without allowing them 
to spread apart. All this beam optics 
requires careful designing of magnetic fields, with very precise spatial 
distribution. Thus an accelerator physicist has to deliver a beam of particles 
accelerated to speeds close to that of light\footnote{for the case of 1 
TeV protons mentioned above $v = 0.9999995 \times c$}, containing a large 
number of particles ($\sim 10^{11}$  or more for the LHC at the nominal 
design) and with a small transverse size (a few $\mu m$ at the collision 
point for the LHC). The beam has to maintain this size while the particles 
are transported across the long periphery of the machine ($27 km$ for the 
LHC ring), many times ($\sim 10^6$) over. Clearly this whole exercise poses 
extreme technological challenges and requires fine engineering.\footnote{Note 
that it is not enough to make these beams of particles of high energy. It has 
to be accompanied by the construction of equally intricate detectors to `detect'
traces of particles produced in these collisions.}  

The decisions about the type of particles which should collide and the 
energies to which they should be accelerated, are  guided crucially by our 
current theoretical understanding of fundamental particles and the interactions
among them. This interplay of different sub-disciplines in the field of high 
energy physics is seen in  the citation of the  1984  Nobel prize for physics 
which was shared by the experimental physicist Carlo Rubbia and the 
accelerator physicist 
Simon Van Der Meer ``for their decisive contributions to the large project, 
which led to the discovery of the field particles W and Z, communicators of 
weak interaction".  The large project mentioned in the citation was the Super 
proton-antiproton Synchrotron: the $S p \bar p S$  which started operation 
at  CERN in Geneva, Switzerland in 1983. This project involved converting
the  proton-proton collider which had been commissioned in 1976, into a 
proton-antiproton collider. The need to do so was indicated by theorists' 
calculations which showed that given the limited energy to which the proton 
could be accelerated using the then available technology, the feat of producing
the $W/Z$ bosons in the laboratory, could be achieved only if one collided 
protons on {\it antiprotons} and that too in large numbers. This prediction 
was made using the then established Glashow-Weinberg-Salam model (put forward 
in 1968) which gave a unified description of the electromagnetic and weak 
interactions\footnote{The inventors
of this theory called the EW theory, had already been awarded the Nobel prize
in 1979.} and with the knowledge of the momentum fraction of the proton 
carried by some of its  constituents, the quarks and the 
anti quarks\footnote{Friedman, Kendall and Taylor, who in 1968 performed the 
first experiment which gave indication of the quark structure of the proton, 
received Nobel prize for physics in 1990.}. Simon Van Der Meer's discovery of `stochastic cooling' made 
it possible to produce tightly focused antiproton beams whereas Rubbia's 
vision and drive made it possible for the project to be realised. 
Already this little discussion tells a lot of things about the current state 
of play in high energy physics, including the need for `project' leaders to 
take these big projects from conception to realisation and the very high level 
of collaboration necessary, not just among the large number of experimentalists
working together on one single experiment, not just among them and the 
theorists, but also among these two groups and the accelerator physicists. The 
particle physics community has been grappling for decades and with reasonable
success, with running these mammoth collaborative projects and at the same time
keep individuality, the basis of all progress, alive!

In this narration, I will first try to explain why LHC: the machine and the 
experiments with it, are a watershed for particle physics. I will then sketch
the essential link between the development of particle accelerators and that 
in our knowledge of the laws governing the interactions among the fundamental 
particles, showing how  this journey has reached destination LHC. After this I 
briefly describe how the high energy physics experiments of past few decades 
have provided important pointers to the physicists who are hunting for the 
Higgs boson and other new particles/physics at the LHC. This discussion will 
then help us understand  how the decisions for the LHC design; the energy and
number of particles in the beam, were arrived at. I will then end by telling
about the LHC physics agenda and the time line in which the particle physicists 
hope to achieve it. 

\section{The story of LHC}
\subsection{LHC: a watershed experiment}

As we go along we will try to understand, why particle physicists believe that,
at least one, as yet undiscovered particle, must exist out there and the LHC 
will see it. Independent of whether this turns out to be the Higgs boson with 
the properties predicted by the corresponding theory, we also expect existence 
of even more particles/interactions which LHC will be  in a good position to 
find should they exist.
One thing is for sure.  LHC : the machine and the experiment, is going to be 
a watershed for the subject of  fundamental particle physics. Particle
Physics will never be the same, independent of what the experiments find.
LHC is now the doorstep and the day of reckoning at hand. We expect the  
LHC to find the last missing piece of the standard model, the Higgs boson.
We expect much more. We believe LHC will also point the way ahead (or even to
a dead end?), help  unravel the deepest secrets of nature and space time.
Hence  it is a watershed experiment. 

To appreciate why this is so, 
it is important to understand the current state of play in the subject. 
The normal course through which science progresses is well known to all of us.
Barring the work of geniuses like Einstein, normally, existing theory 
and observed phenomena which are unexplained in the framework of that 
theory, lead to new theoretical developments. This then leads to 
predictions, which then get tested in experiments. However, in the subject
of High Energy Physics (HEP) one is in a very strange situation. We have a
theory, called Standard Model (SM) of particle physics, which works so well
that there seems to be almost no unexplained phenomena.(See Box-I, for some
details of the SM.) Various new theoretical developments have taken place 
only by the demands made by the community on the properties that the 
mathematical theory aught to have for it to be a 
satisfactory description of the fundamental interactions. One example of a very
simple demand is called unitarity: which very simplistically stated means
that probabilities for different events are  bounded by unity, as they always
must be. Some of the theoretical developments beyond the SM (BSM) have also
come by demanding that some features, such as the values of the masses that 
different particles have, should be understood from first principles in the 
theory. For example, in the SM, we have such a first principle understanding 
why the rest mass of the photon is zero. All the efforts to find {\it direct}  
proof for any of these new ideas (Supersymmetry, additional compact dimensions 
of space\cite{bsm} to name a few), in the various particle physics 
experiments and/or to look for their implications in cosmology/astrophysical 
situations, have yielded, so far, negative results.
In other words, we seem to have found a `perfect' theoretical description
of fundamental constituents of all the matter and interactions among them,
barring a `direct' verification of one last `piece' of the puzzle, the Higgs 
boson.  

On the one hand, the particle physics community has  strong theoretical 
reasons that there is new physics at the TeV scale, while on the other, the 
experimental evidence for its `need' is only in the form of tantalising 
`indications' such as  non-zero masses for the neutrinos, or the
dark-matter in the Universe etc. It should be emphasized here, that the 
track record of particle physicists is pretty good so far in this context and 
theoretical developments based on demands of  aesthetics  alone have been 
fruitful at getting at the root of some of the very fundamental questions 
about laws of nature. But, admittedly, the time gap between theory and 
experiment has never been so big as it is at present. We are still awaiting
the final experimental verification of a theoretical advancement made in 1964.

One direction in which progress can come is 
by increasing the available energy at which particle interactions are studied.
There are solid reasons to believe that experiments at an energy  of the 
order of a few Terra electron volts (TeV) would bring further progress.
The HEP community expects the three TeV colliders (the $p \bar p$ Tevatron
in USA, the $pp$ collider LHC which has started its journey now and the 
International Linear electron-positron Collider (ILC) which is now under 
planning) to help us see the way ahead. Due to the higher energy of the LHC 
than the Tevatron, at present all the particle physicists look to the LHC to 
provide the final clinching evidence for the SM and give at least a glimpse 
of the physics beyond the SM (BSM), which all of them  believe, must exist at 
around the TeV scale.

\subsection{Cosmic Connections}
Very interestingly the fundamental laws of particle interactions operating at 
the `heart of the matter' at `'femto' meter scales (or smaller) that the particle
physicists have been studying, have implications for issues 
cosmological in nature, concerning the universe, as it was at the
beginning and what it is now. The formation of protons/neutrons in the early
Universe (Nucleosynthesis), the observed abundances of various elements such
as $He,Li$ etc. in the Universe, can now be understood in terms of known 
physics of these interactions 
and experiments performed in terrestrial environment and laboratories.
However, it is clear now that there are some very basic features of our
universe that seem to indicate existence of particles outside the list 
given in the Tables 1 and 2 of Box-I. Among these are the following 
facts:
1) The fundamental particles listed in the Tables 1,2  of Box-I constitute only 
$4 \%$ of the total visible matter in the Universe and $23 \%$ of the matter in
the Universe that is is `invisible'(so called Dark Matter,DM) does not consist 
of particles of the SM, 2)Among the particle content of the Universe, we see 
only matter particles and virtually no antimatter particles. To be precise the 
difference between the number of baryons and anti-baryons normalised to the 
total number of photons, $\frac{N_B - N{\bar B}}{N_\gamma} \sim 10^{-7}$,
\comment{\footnote{If that were not the case, we would have seen results
of annihilation of matter-antimatter, happening in the universe.}}
and 3)The universe seems to be continuously accelerating. The source of this 
acceleration is again not found among the known matter/interactions and seems
to indicate an unknown form of `energy' (hence called 'dark energy') which 
forms about $73 \%$ of the total mass/energy content of the Universe. All 
these indicate very clearly existence of physics beyond what is in the SM.

What is even more interesting is that possible candidates which can throw light
and/or provide solution to these issues, also exist very `naturally' in almost 
all the ideas of BSM physics that particle physicists have putting forward 
forward for an entirely different reason, viz., to remove some of the 
theoretical shortcomings of the currently accepted description of the SM.
But no one idea separates itself completely from the crowd. Experiments
are the ultimate Jury which would choose between these different ideas. At the 
LHC, in addition to hunting for the Higgs boson, the last piece of the SM 
puzzle, it may be possible to create some of the particles which make up
about $23 \%$ of the total matter in the universe, the dark matter that is
`invisible' to light; or probe the physics that makes the universe of today 
contain much more  `matter' than anti-matter. 
The LHC can probe this synergy between the world at the femtoscale and 
the physics of the entire universe and provide answers to some very basic
question about nature of things! This should clearly show that the `stakes' 
in these collider experiments are really high!!

\subsection{Accelerators \& particle colliders: journey unto LHC}
To understand the unique role particle physicists expect the LHC to play,
it is a good idea to briefly look into the history of particle accelerators
These high energy particle accelerators are like microscopes which have 
allowed us to peer at the 'heart of matter'.  High energy proton and electron 
beams produced using these accelerators have been used in two different modes:
1) The Fixed Target Machines where $e, \mu, \nu$ and $p, \pi$ beams 
are incident on a stationary target which consists of light or heavy nuclei.   
2) Colliders where beams of accelerated particles collide against each other. 
In the latter  class only the $e^+,e^-, p$ and antiproton beams can be
made to collide in enough numbers to make the experiments meaningful.

In case of a fixed target machine, for a beam of energy $E_b$ incident on a 
target of mass $M_T$, total energy available for new particle production is 
$E_{cm} = \sqrt{M_T E_b}c$, whereas in the collider environment, specialising 
to the case where both the beams have particles with same mass and energy, the 
energy available for particle production is $2 E_b$. Recall that mass of a 
proton, a typical target, is $\sim 1 GeV/c^2$. Thus the collider mode is 
superior for new particle production than the fixed target mode, from the point
of view of energetics, as beam energies approach $\sim $ GeV. For example, 
a particle called $J/\Psi$ with mass $3.1 GeV/c^2$, now understood to be a 
bound state of a charm-quark and an anti-charm quark, was produced for the 
first time in 1974 in an $e^+e^-$ collider called SPEAR at SLAC with a 
beam energy of $2$ GeV, whereas soon after, the same particle was created in 
a fixed target experiment at the Brookhaven National Laboratory which 
employed proton beams accelerated to $30$ GeV.

But energy is only one consideration. Equally important are the Luminosity 
$\cal{L}$ of collisions, i.e. the number of collisions per unit area, per 
unit time and the kind of interactions that the colliding particles possess. 
Colliders became a popular tool only after the accelerator physicists developed
better techniques to make intense, well focused beams. In fact the SPEAR 
collider mentioned above, was one of the early example of a collider 
experiment. Till then the burden of progress was carried mainly by the fixed 
target experiments. 

Through the early part of these explorations,
experiments with beams of higher and higher energy just revealed constituents 
at smaller and smaller distance scales. After the discovery of the  quarks, 
in 1968, lying at the heart of protons and neutrons, the later increase in 
energy has brought about production of the force carriers and help develop/test
the theory which can describe the interactions among the fundamental 
constituents.  Experiments with different  fixed target machines and the 
colliders together, provided this information.

The colliders which helped particle physicists in this journey can be  
divided in to three different classes based on colliding 
particles. The leptonic colliders: 1)electron-positron ($e^+e^-$) and 
2) electron-proton ($e^- p$) colliders and the hadronic 
colliders: 3)proton-proton($pp$), proton-antiproton ($p\bar p$) ones.  

The story began with $e^+e^-$ circular machines in the 60's at Frascati(Italy),
Novosibirsk(Russia) and Orsay(France). It then proceeded through SPEAR (1973), 
PEP (1980), SLC(1990) at SLAC, Stanford in USA; DORIS(1973),PETRA(1979) at
DESY, Hamburg in Germany; LEP-I(II)(1989) in CERN, Geneva, Switzerland,
with the  beam energy increasing from a 1.5 GeV at SPEAR to up to (15) 23.4  
GeV at  PEP (PETRA), 30 GeV for Tristan, 45 GeV for LEP-I, SLC and 104.5 GeV 
for LEP-II. The SLC was the only 'linear' collider among all these.\footnote{
BELLE (KEK: Japan) and BABAR (SLAC: Stanford) are the higher 
intensity but lower energy, $E_b \sim 4.5 GeV$ machines, carrying
on the legacy of DORIS at DESY in Germany and CESR at Cornell in USA. This 
concerns travel on another front in experimental HEP and not discussed further 
here.} 
Finally the story 
continues now to a possible International Linear Collider (ILC) or Compact 
Linear Collider (CLIC). It is not yet clear if and where the last mentioned 
colliders will be built but feasibility studies for these two already
go on. 

Next colliders to come into play were the proton-proton (pp) and 
proton-antiproton ($p \bar p$) machines. The PS (proton synchrotron) built more
than 50 years ago at CERN (and still working today) fed high energy proton beams
to the Intersecting Storage Ring (ISR) : world's first $pp$ collider (1971). The
story then moved on to the Super Proton Synchrotron to achieve $ \bar p p$ 
collisions at  the  $S p \bar p S$ collider at CERN, in 1983.  Close on its 
heels was the Tevatron in USA and now the hadronic machine at the forefront is 
the $pp$ machine is the LHC.

Interspersed with these pure leptonic and hadronic colliders, was a machine
which collided  electrons/positrons on  proton. This $ep$ collider HERA in 
Hamburg, Germany, provided invaluable information on the quark and gluon
content of the proton, which as we will see below is required to predict 
what particles the LHC can produce and at what rate.  

A list of some of these different leptonic and hadronic colliders,
of relevance to the discussion here, is presented in the tables~\ref{tab3} 
and \ref{tab4}.  Due to the differences in the nature of the hadronic and
leptonic colliders, they have played very complementary roles in our quest 
for the fundamental constituents of matter and interactions among them. 
\begin{table}
\begin{center}
\begin{tabular}{|c|c|c|c|}
\hline
&&&\\
Period&Type&Energy GeV&Perimeter\\
&&&\\
\hline
1971-1976&pp (ISR) Circular,CERN& $32 \times 32$ & $\sim 7 km$\\
\hline
1983-1985&$p \bar p$ $S p \bar p S$ , Circular,CERN&$270 \times 270$&   "\\
\hline
1987--&$p \bar p$, Circular, Tevatron, USA &{\bf $980 \times 980$}&$\sim 6 km$ \\
\hline
2009&LHC, $p p$,Circular&$1180 \times 1180$ &$\sim 27 km$\\
2010--&CERN&{\bf $3500 \times 3500$} &\\
\hline
\end{tabular}
\end{center}
\caption{List of hadronic colliders of interest, two are still in action.}
\label{tab3}
\end{table}
\begin{table}
\begin{center}
\begin{tabular}{|c|c|c|c|}
\hline
Period&Type&Energy GeV&Perimeter (Circular)\\
&&&Length (Linear)\\
\hline
1973-1983&$e^+ e^-$, Circular&$1.5\times 1.5 $& $\sim 0.6 km$\\
&SPEAR, USA& $3.5 \times 3.5$&\\
\hline
1978-1986&$e^+ e^-$, Circular&$6.0 \times 6.0$&$\sim 2.3 km$\\
&PETRA, Germany&$23.4 \times 23.4$ &\\
\hline
1990-2007&$e^{\pm} p$, Circular&$26.5 \times 800$&$\sim 6.3 km$\\
&HERA, Germany&&\\
\hline
1989-2000&$e^+ e^-$, Circular,CERN& $\sim 45 \times \sim 45$&$\sim 27 km$\\
&LEP-I, LEP-II & $104.5 \times 104.5$&\\
\hline
1989-1999&$e^+ e^-$, Linear& $50 \times 50$ &$\sim 3.2 km$\\
&SLC, USA&&\\
\hline
???&$e^+e^-$,ILC,Linear&$500\times 500$&30 km\\
???& $e^+e^-$, CLIC,Linear&$1500 \times 1500$&$\sim$ 20-40 km\\
\hline
\end{tabular}
\end{center}
\caption{Some of the Leptonic Colliders : past and in planning.} 
\label{tab4}
\end{table}
The hadronic and leptonic colliders have different advantages and disadvantages
For $e^+e^-$  colliders the initial beam energy is very accurately known 
as the colliding particles are the same ones which are being accelerated.
For the $pp$ or $p \bar p$ machines the colliding particles are
composites and at high energies the colliding fundamental particles
are the (anti-)quarks, gluons: the partons.  From the measurements at HERA
and other fixed target machines, it is known that, on the average only 1/6 th  
energy of the proton is available to the colliding partons.  Thus for the same 
energy of the beams, $E_{cm} (e^+e^-) \sim   6 E_{cm} (pp)$. Thus to probe 
physics at a given energy scale, one needs $pp/p\bar p$ colliders, with higher 
energies than the $e^+e^-$ machines. But then it {\it is} easier to accelerate 
the $p/\bar p$ to much higher energies. Further for a given beam energies,
these hadronic machines can provide a broad range of energies at which 
collisions between partons can happen. Thus a hadronic machine allows us 
a broad sweep of energies and one can take a quick/dirty  look at the 
landscape, as opposed to an $e^+e^-$ machine where for a fixed beam energy
the collision energy is fixed too! Hence, traditionally the hadronic machines
have been 'discovery' machines and leptonic colliders have been the precision
measurement machines. Further, we really need a hadronic machine to study
processes initiated by strongly interacting particles: quarks and gluons.

The physics flow among these different machines is very interesting. For 
example, the studies of $ep$ collisions at the Stanford Linear Accelerator(SLAC)
revealed(1968) that the proton is made up of quarks. Further experiments,
in the next three decades with $\mu p$, $\nu p$ collisions from  
Tevatron/CERN and from $ep$ collider HERA, revealed  that the proton contains 
 quarks,anti-quarks and gluons (all three collectively called partons);
the evidence for the latter being `indirect'. These experiments further gave 
extensive  information on proton momentum fractions carried by the partons. 
The $pp$ collider ISR at CERN gave the first evidence of the parton structure 
of the proton in $pp$ collisions in 1973. 

The experimental evidence for the  first of the heavy quarks and leptons, 
the $c$ quark and $\tau$ lepton, happened at the $e^+e^-$ collider SPEAR and 
the fixed target experiment at Brookhaven, in 1974-1975. Both, the existence 
and the mass of the charm quark was predicted in 1970 based on the EW model of 
1964, mentioned earlier. The $\tau$ lepton, discovered at SPEAR, came 
uninvited to the party. However, its presence was soon found to be necessary, 
from  theoretical considerations, in order to match the pair of quarks whose 
presence was predicted from an analysis of the well established phenomenon of 
CP violation. The two quarks, so predicted, were bottom/beauty(b) and top(t).
Out of which the bottom was discovered in a fixed target experiment in 1977, 
but the discovery of the top quark at the Tevatron had to wait till much later.  
The `direct' observation of gluons at the $e^+e^-$ collider 
PETRA in Hamburg, Germany in 1978, was then followed by the direct observation 
of $W$ and $Z$ at the CERN $p \bar p$ collider in 1983. Their mass had been 
predicted while planning this collider, using the results from the experiments 
at the lower energy  $e^+e^-$  colliders like PEP/PETRA and TRISTAN along with
those from the fixed target experiments using $\nu$ beams. The first production
of a handful (in two digits) of $W,Z$ at UA-1/UA-2 at CERN in 1983 was then 
followed by production of {\it millions} of $Z$ at LEP and SLC in 1990-2000.

This is pattern that is repeated time and again. 
The $pp$ or $p \bar p$ machine which is a higher energy, broad band machine 
makes discoveries, whereas the cleaner environment of $e^+e^-$ colliders 
allows for precision studies. Thus the lepton and hadron colliders have been 
alternately taking the  role of leading machine, driving the progress of 
particle physics. The precision measurements at LEP-I and LEP-II, having
been  complemented by the hadronic collider Tevatron, which discovered the
last missing quark, the top quark, now time has come for a higher energy broad 
band machine. 

The information on the main physics discoveries made at different colliders
have been 
\begin{figure}
\begin{center}
\includegraphics*[width=12cm,height=9cm]{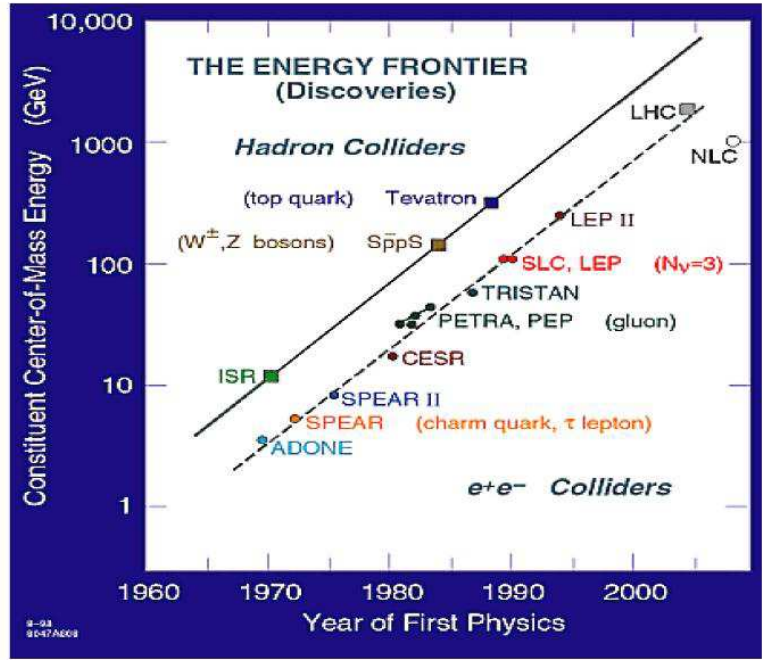}
\end{center}
\caption{A summary of particle physics discoveries made with different machines}
\label{fig:phys-acc}
\end{figure}
summarised in Fig.~\ref{fig:phys-acc}. This shows how the baton for 
discoveries has moved between different machines and how the energy frontier 
has moved. 
Missing from this figure is the fixed target experiment at SLAC, with electron
beams of energy up to $50$ GeV which made the discovery of light quarks (u,d)
inside proton  in 1968 and the  fixed target machines which followed it at 
Fermilab and CERN, with $\mu,\nu$ beams up to energy 800 GeV, which helped 
confirm the existence of the strange quark (s).

This shows that the colliders and the fixed target machines have functioned
in tandem, revealing layers of substructure of matter and providing 
information which can be used to elucidate the  mathematical description 
of interactions  among the fundamental particles. With the LHC, experimental 
data using the higher energies reached by the LHC can push knowledge forward, 
challenging those who seek confirmation of established knowledge, and those 
who dare to dream beyond the paradigm.

\subsection{Precision testing of the SM and pointers to the Higgs hunters}
The most important part of the intellectual achievements in the area of 
particle physics over the past five-six decades, arrived at by using the 
information from these accelerators and colliders has been the understanding 
of the three fundamental interactions (shown in Table 2 of Box-I), in terms of 
exchange of the force carriers among the matter particles.
Just like Quantum Mechanics is the correct mathematical framework to describe 
the phenomena that occur at short distance scales and/or high energies, the 
mathematical framework which can be used to describe and calculate 
processes involving creation and annihilation of particles is called Quantum 
Field Theory (QFT). In fact the description of all the three interactions,
strong, electromagnetic and weak, in terms of QFT's with special properties,
is the basis of the SM of particle physics. At present a complete 
mathematical description of all -- the low energy and high energy-- phenomena 
in the world of fundamental particles, is possible in this framework.\footnote{
A lot of these achievements were recognised by a large number of Nobel prizes 
in Physics in the past fifty years, beginning with the Nobel prize to Feynman,
Tomanaga and Schwinger in 1960 and ending with the Nobel prize to Y. Nambu, 
M. Kobayashi and T. Maskawa in 2008.} To get a feel for the level of precision i
in the experimental measurements and that in the theoretical predictions, 
I present in Table~\ref{tab5}, details of some of the most crucial parameters 
of the unified theory of electromagnetic and weak interactions (EW theory), 
which I have taken from Refs.~\cite{lepewwg,PDG}.
\begin{table}
\begin{center}
\begin{tabular}{|c|c|c|}
\hline
Observable&Experimentally measured value&SM fit\\
\hline
Width of the $Z$ boson: $\Gamma_Z $&$2.4952 \pm 0.0023$ GeV& $2.4959$ GeV \\
\hline
Mass of the $W$ boson: $M_W$ &$80.404 \pm 0.030$  GeV$/c^2$  & $80.376$ GeV$/c^2$ \\
\hline
Mass of the $t$ quark $M_t$ & $172.5 \pm 2.3$ GeV$/c^2$ & $172.9$ GeV$/c^2$\\
\hline
\end{tabular}
\caption{Precision testing of the SM}
\label{tab5}
\end{center}
\end{table}
The level of precision of the EW theory predictions as well as the experimental
measurements and the agreement between the two is quite impressive
\footnote{In fact the almost complete 
success of the SM, has led some theorists to wonder whether time has come for 
a paradigm shift. Further, a quantum theory of the fourth fundamental 
interaction, the gravitation, seems to lie outside the realm of QFT's 
describing the three interactions of the SM, the Gauge  Field 
theories. String theorists believe that Quantum Field Theories are  sort of a  
just a 'low energy' paradigm and string theories might be the language
to  use once you want to include gravitation. Jury is more than out on
this point. It is possible that the LHC might tell us that this belief 
of the string theorists is the truth! Then again it may not!}.

However, it has to be said that in order that the theoretical computations 
leading to theory predictions in Table~\ref{tab5} can be performed, in addition
to the particles listed in Tables 1,2 of Box-I, one more particle has to exist
in the SM. The so called Higgs boson with spin 0, i.e.a particle 
with no spin degree of freedom. The SM has precise predictions for all the 
interactions of such a particle with all the other particles. Theoretical 
predictions for various EW observables do depend on the mass of the Higgs 
Boson, $M_H$. The SM predictions shown in Table~\ref{tab5} are calculated 
using  a particular value for $M_H$. 
Hence, these  precision measurements can in fact be used to derive and 
`indirect experimental' upper bound on $M_H$. This can be then further 
combined  with the experimental lower limit from the non-observation of the 
Higgs boson in all the experiments (at LEP as well as at the Tevatron) up to 
now. The range of $M_H$ allowed by the current experiments, at $95 \%$ c.l.,
in the SM is $115 < M_H c^2  < 150$ GeV~\cite{Moenig}. These limits are of the 
same
order as $M_W/(M_Z) \simeq 84/(91) GeV/c^2$.  In particle physics parlance one says 
that the experiments prefer a 'light' Higgs boson.
Without going into technicalities, let me also mention that in the SM the
mass of the Higgs boson is not predicted. But, pure theoretical considerations, 
making no reference to the above mentioned precision measurements, predict lower 
and upper limit on its mass. 

Thus information available from earlier colliders in fact set the goals for
the LHC just like Gamow's theory predicted goal posts for energy required
for radioactivity. The physics flow between different machines  and their
implications for the LHC are summarised
\begin{figure}
\begin{center}
\includegraphics*[scale=0.6]{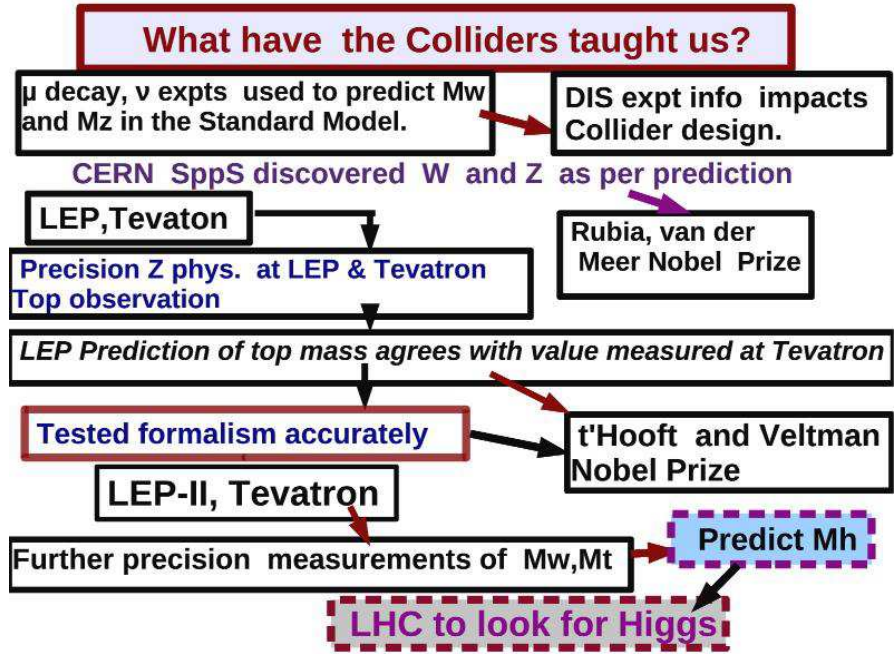}
\end{center}
\caption{The flow of physics information between different colliders
and their impact on design considerations of the colliders.}
\label{fig:accmh}
\end{figure}
in Fig.~\ref{fig:accmh}. Thus we see that results at the current series of
machines have always driven the physics at the next generation of machines,
giving an indication of the required energy/luminosity. The framework of the
SM was abstracted using results from the fixed target $\nu$ experiments and
from the low energy $e^+e^-$ colliders, DORIS/PEP/PETRA/TRISTAN. In that 
framework one had a prediction for the masses of $W/Z$ bosons. The $UA$--$1$ 
and $UA$--$2$ experiments (UA standing for Underground Area) measured the
masses of these $W/Z$ bosons producing them directly, thus giving the first 
`direct' confirmation of the SM predictions of these masses.
The combined measurements at the LEP-Tevatron, combined with high precision 
theoretical calculations in the framework of the SM, then led to a prediction 
of the top mass. The agreement of this predicted value of $M_t$ with the one 
directly determined at the Tevatron (Table~\ref{tab5}), proved the correctness 
of the SM to a high degree of accuracy. But these highly precise calculations 
are possible in the
SM only if the particle spectrum of Tables 1 \& 2, is complemented by a Higgs 
boson. These precision measurements and calculations, now predict the range 
in which the  $M_H$ value must lie, in the SM. Further the theory itself 
predicts that $M_H$ should lie in a certain range, SM or BSM, based on purely 
theoretical arguments. To cover this entire mass range, then a high energy, 
broad band hadronic machine is needed. The Tevatron does not have enough 
energy for this purpose. This thus makes the case for the LHC. Of course 
this will need to be followed by a TeV energy $e^+e^-$ collider, which is 
already being planned as the LHC is running\footnote{This too is quite common 
in HEP.  LHC planning was started, when LEP was in construction and the size 
of the LEP tunnel was decided by keeping in mind that one would want to build 
a LHC someday in the same tunnel.}.

\subsection{LHC design}
If the SM is correct a light Higgs (ie. with mass comparable to that 
of the $W/Z$) {\bf must} be found experimentally.  There is strong evidence
that the SM is a very good approximation to reality. Right now  LHC is the 
only collider which will be able to find a 'light' Higgs boson\footnote{The 
Tevatron too can look for a 'light' Higgs but is not so effective in that
mass range and it is also not clear how much longer it will run.}.  
\comment{
One should mention that both the experimental and theoretical limits 
on the Higgs boson mass are different depending on whether there is 
physics beyond the SM and what is the energy scale of this BSM physics.}  
Even if the SM is not the entire story and hence the Higgs is not in the 
low-mass range predicted by the SM or is not present at all, the success of
the EW theory in explaining the precision measurements indicate that 
a `look-alike' of the Higgs boson must be present. The general theoretical 
bounds on $M_H$ mentioned earlier, encompass those that one computes for the 
`look-alike's as well. Hence, observation of the Higgs, measurement of its 
mass or even the non observation in a given mass range, will shed light on the 
puzzle of the formulation of a unified theory of electromagnetic and weak 
interaction (EW theory) and the apparent mystery of the breaking of the
corresponding symmetry at low energies\cite{hinsa}.

It was clear while planning the LHC that one has to design the machine such 
that it should be able to probe the issue in a completely model 
independent fashion. A very general theoretical upper limit on the Higgs mass 
is  about $900$ GeV. Even if the SM is not the complete reality, hence no 
light Higgs boson is found, it would  still be possible to unravel the mystery 
of the EW symmetry breaking by studying effective WW scattering at a total 
energy of 1 TeV. The choice of energy and luminosity of the LHC was made by 
by demanding that LHC should be able to cover this eventuality.\footnote{This 
choice could be made even before the results of the precision measurements at 
LEP were available to us, because of the general nature of the argument.}

It can be understood somewhat simply as follows. 
\begin{figure}[hbt]
\begin{center}
\includegraphics*[scale=0.5]{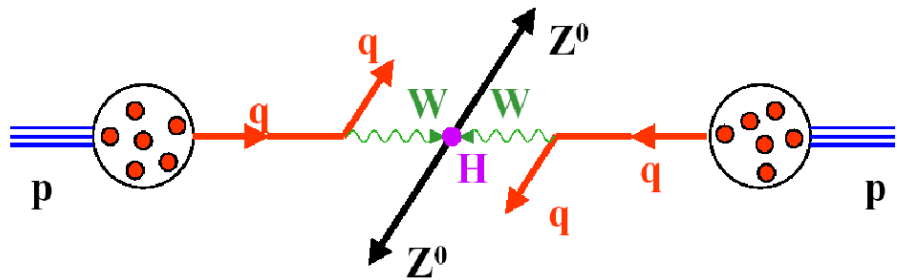}
\end{center}
\caption{How LHC parameters were decided}
\label{fig:lhcdesign}
\end{figure}
The available knowledge of the quark/gluon content of the proton, indicated
that unlike the case of the earlier CERN and FNAL colliders, the physics 
potential of the machine would be independent of whether one has a $pp$ 
machine or a $p \bar p$ machine. Hence a decision to make the, cheaper and the 
easier to build , $pp$ machine was taken. Thus the LHC collides proton on 
protons. As said earlier, the protons are made of quarks and gluons. Hence
the collisions at LHC, are effectively collisions among these quarks and
gluons. They carry only a fraction of the energy of the proton. 
Fig.~\ref{fig:lhcdesign} shows a possible scattering process
$WW \rightarrow Z Z$. For the two W's to have a total energy of 1 TeV, each of
them must have energy of $0.5$ TeV and hence  each of the parent quark in the 
figure must have energy of 1 TeV. From the measured distribution of the momenta
of a proton among its quarks, one knows that on the average, the quarks carry 
about 1/6 th energy of the proton. This means that the protons must have an 
energy of 6 TeV. Hence one planned on a $pp$ collision energy of 7 TeV on 7 
TeV. The total number of $pp$ collisions were decided by using the theoretical 
estimate for the rate of production of the events 
$pp \rightarrow WW \rightarrow ZZ $  and demanding that  at least $10$ events 
per year be produced. Note how our theoretical knowledge about the SM, has 
set the bar for the new machine energy and intensity. 

\section{LHC agenda}
\subsection{LHC agenda:SM}
Thus item number 1 on the LHC agenda is to find the Higgs boson and measure its
properties, such as mass and the interaction. The way the LHC has been designed,
at  its planned energy and luminosity, it will be able to search for the Higgs 
boson over the entire mass range, allowed on very general principles, 
irrespective of whether the SM is the whole story or otherwise. Due to the very 
nature of the hadronic machine, LHC will be able to do the second job of 
measuring its properties only to a  moderate accuracy of about $10 -20 \%$. A 
more accurate measurement has to wait for the next generation $e^+e^-$ 
collider which will have less energy than the LHC, but the measurements will be
much cleaner and more importantly the LHC would have indicated to us the Higgs 
mass and hence the optimal energy to run the machine. 

I have focused here on what I believe to be the main reason for planning and 
executing this mammoth project. Interestingly, however, it transpires that the
LHC will also offer almost a direct  probe into what went on in the early 
Universe, in addition to the indirect probes that I have talked about. This 
will happen in the `heavy ion avatar'  of the LHC, where instead of protons, 
relativistic heavy ions will be colliding against each other. These collisions 
will recreate energy densities that exist in the early Universe. In the 
presently accepted description of the early Universe in the Standard Model of 
Cosmology, the early universe is a hot, radiation dominated plasma, which is 
an 'almost' an ideal gas.
\begin{figure}
\begin{center}
\includegraphics*[scale=0.5]{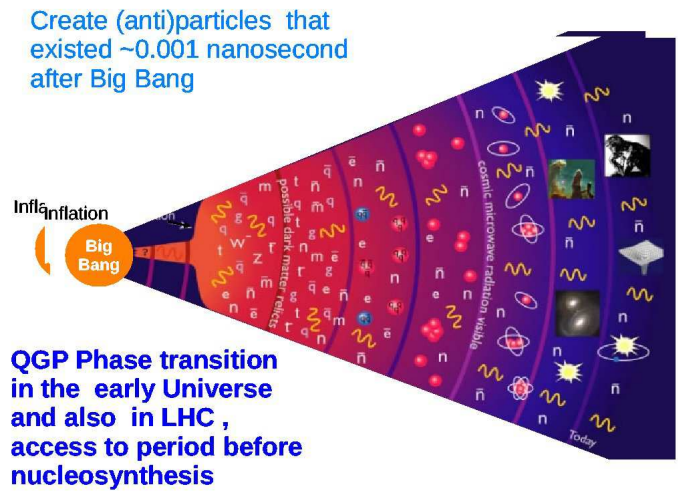}
\end{center}
\caption{Depiction of different stages in the evolution of the Universe}
\label{fig:qgp}
\end{figure}
This picture, shown in Fig.~\ref{fig:qgp}  has been verified up to a temperature
of  $\sim$ 1 MeV, i.e. the Nucleosynthesis period as mentioned earlier. 
The period before that
is opaque to current astrophysical measurements. In Heavy Ion Collision option
one can recreate the transition from the state of a quark-gluon plasma to the 
currently known hadrons. Thus the LHC can create a `mini-bang' as opposed to 
the big-bang which started off our Universe. Study of this transition will 
also offer a look into the strong interaction dynamics in a region that so
far has not been accessible to our experiments and which also poses challenges
to the theoretical computation techniques.

The next question to ask is, will it be enough for the particle
physicists if a  Higgs boson were to be discovered at the LHC with precisely
these properties, in precisely this mass range? Will that conclusively 
complete our understanding of the fundamental interactions among fundamental
constituents of all matter? In other words, what is the need for going beyond
the standard model, the BSM physics, that I have been mentioning.

\subsection{BSM physics}
There are a variety of reasons which have prompted particle physicists to look
for ideas beyond the SM. Majority of these reasons are aesthetically. But a lot
of progress in particle physics (and in fact theoretical physics) has come from
looking for elegant explanation of observed physical phenomena and properties.
For example, $1/r$ dependence of the Coulomb Potential is due to the 
'zero' rest mass of the photon and zero rest mass of the photon is due to 
the fact that Maxwell's equations have gauge invariance. Thus in this case,
the principle of gauge invariance 'explains' the experimentally observed nature
of Coulomb Potential and zero rest mass of the photon. Similarly, as described
in the earlier subsection, the precision testing of the SM, implies that  
a Higgs OR a look alike must exist and data tell us it must be light! 
Just like gauge invariance gave an explanation of the power of $r$ in Coulomb's
Law, we would like to understand why the Higgs is light!! Particularly so
because, a quantum field theory of spin 0 particles would predict that 
the Higgs should be as heavy as can be!!!\footnote
{If one likes one can interpret the direct/indirect experimental indications for
a `small' higgs boson mass  as a 'disagreement' with the SM.}
This is one of the `theoretical' reason for expecting BSM physics.

Among the experimental hints for BSM, the most significant is the existence
of `invisible' Dark Matter' (DM) in the universe. Since, this DM can not
consist of the particles of the SM,  it is a clear indication for
BSM physics. It is a matter of enormous interest, that almost all the BSM
physics scenarios, `naturally' have a particle that could be the DM in the
Universe.  

Second experimental hint of BSM physics is the nonzero, tiny mass
of the neutrinos. It is now firmly established that neutrinos have tiny,
nonzero masses {\it and} the masses of the fermions have a huge hierarchy.
For example the mass of neutrinos is smaller than $1 eV/c^2$ and mass of
the top quark is $\sim 175 \times 10^9 eV/c^2$.
In the SM all these masses are just arbitrary parameters.  A natural question
that has been asked by particle theorists, is whether we can have a fundamental 
understanding of why they have the values they have? Such questions might 
sound esoteric, but a lot of progress in science actually has come from asking 
such questions. Neutrino masses are especially tiny and being neutral, 
it is possible to have elegant mechanism for generating these `small'
masses almost `naturally' in some BSM formulations. 

Further, unification 
of all interactions into one master interaction is a dream which even
Einstein had had. At first level, unification of weak and electromagnetic
interaction in a single Electroweak interaction, has been observed. All the
three interactions do not unify at a given scale if there are no more 
fundamental particles in addition to those in the SM, whereas
such a unification is possible in BSM scenarios. This is one more 
`indication' of BSM physics. It may be added at this point, that almost all 
the models which attempt to explain the observed pattern of fermion masses 
quite often use the framework where the interactions do unify at a high energy 
scale.  Such theories also have mechanisms which can explain the observed 
matter- antimatter asymmetry in the universe or a candidate for dark matter, 
almost automatically.

It is worth observing that that both the experimental 
and theoretical limits on the Higgs boson mass can depend on the (non)existence 
of the BSM physics and its energy scale, if it should exist. Hence observation
of the Higgs and a measurement of its mass can tell us a lot about the energy
scale and nature of BSM physics that might exist. This is like trying to 
decipher the nature of the animal being tracked from its footprints.
A brief discussion of some of the theoretical options for the BSM physics
that have been proposed is given in Box-II.

\subsection{LHC agenda: BSM}
It should be clear from all these discussions that the 
the job of the LHC is not finished even when it finds a light Higgs
boson. If it is indeed in the mass range that the earlier collider
experiments have indicated, the theory  must then explain why it is light! 
Different explanations of why 
the Higgs is light imply new symmetries, new particles exactly at the TeV scale
which should be observed at the LHC.  So if a light Higgs boson is found
the next item one on LHC phenomenology agenda is to infer which one of the
many explanations (if any) is right.  

A natural question is what if we 
don't find it?  It would mean that there is an alternate to EW symmetry 
breaking which passes the challenge posed by the precision tests as 
comprehensively as the SM does.Then the LHC agenda item is to check which
one of these alternates, if any, is correct !  So one of the big area of 
phenomenological research has been how to delineate different BSM ideas 
from each other~\cite{belyaev} at the LHC. 

From the discussions of the different BSM physics scenarios it is clear that 
the hunt for these BSM scenarios may in fact  help us explore further
at the heart of matter and probe the structure of space time. Most of the time
due to the very nature of these BSM ideas, they have implications for the
early Universe and hence this is an opportunity to test some of
the ideas of the Standard Model of Cosmology at the LHC.

I reproduce below a list of objectives that have been set out in a road 
map of particle physics for the next decades by the world community.
\begin{itemize}
\item Are there undiscovered principles of nature:
      New symmetries, new physical laws?
\item Are there extra dimensions of space?
\item  Do all the forces become one?
\item Why are there so many kinds of particles?
\item What is dark matter?
      How can we make it in the laboratory?
\item What are neutrinos telling us?
\item How did the universe come to be?
\item  What happened to the antimatter?
\item How can we solve the mystery of dark energy?
\end{itemize}

Apart from the last point in this list, we expect the LHC to shed light
on almost all the points. That is the reason LHC is the watershed experiment
in Particle Physics. All of us are looking to it to point the way ahead.!

\section{LHC machine and physics: time line}
\subsection{LHC machine: accident and repairs}
Even by the standards of HEP laboratories designing and building the 
Large Hadron Collider was a challenging exercise. As said above, LHC was 
conceived in 1980s and  the planning
began already in 1989. The tunnel which houses the LHC now, was built
between 1983-1988  and was home to the LEP experiment till 2000. CERN Council
approved construction  in 1994. Limiting oneself to the use of the LEP
tunnel meant that there was an upper limit of 7 TeV to the energy to which 
the protons could be accelerated\footnote{To reach the same physics goal of
being able to hunt for the Higgs up to the general upper limit of about 900 GeV,
the then under planning and later canceled, Superconducting Super Collider
(SSC) project in USA, was supposed to accelerate the beams to an energy of
20 TeV in a much bigger tunnel.}. Hence the LHC had to plan on larger 
luminosity. This meant smaller 
bunches, higher magnetic fields and  packing more particles per bunch. The 
LHC builders then decided to use new methods for acceleration. They
also  decided to use a very 
innovative idea where the beams would circulate in two separate rings 
just above one another. For $e^+ e^-$ or $p \bar p$, the same magnetic
field automatically suffices to steer the two bunches in opposite directions.
That is not the case when both colliding particles have the same charge as they
do for the LHC. All this was not just technologically challenging, but also
expensive. The decision was then taken by the CERN Council to build it in two 
stages. In the meanwhile plans for the SSC were abandoned. From  1995 onwards,
countries which were not members of CERN, viz.,  Japan, USA, Canada,
India and Russia, promised support to the LHC machine and decision was 
taken to do it in one go. While the Indian HEP community had participated
in building detectors  and doing experiments at the earlier fixed target and 
collider facilities,  this was the first instance where India participated 
in the building of the machine itself.

Given the ring size and the energy to which particles needed to be accelerated,
meant that one needed to have higher magnetic fields of about $8$ Tesla. This 
in turn  meant that the magnets had to be cooled down to $1.9^{\circ}$ K. 
Contrast this with the other superconducting collider Tevatron where this 
temperature is about $5.2^{\circ} K$ and the magnetic fields required about a 
factor two smaller. The magnetic fields required at the LHC are also higher 
than the one used at the old SPS, by about a factor 5. Not just that, 10-12 
Tesla is about the upper limit at which these Niobium-Titanium accelerator 
magnets can function, still remaining superconducting. All this should give a 
flavour of the engineering and technological improvements that were required 
for the LHC. The most crucial piece of the machinery for the LHC are the 1232 
dipole magnets, each weighing 34000 Tonnes and costing about 0.5M CHF each.

Before the start up of September 2008, the energy goal was already lowered from
the initial $7$ TeV to $5$ TeV, with plans to raise the energy afterwords. Its
start up in September 2008, came to an abrupt end during a test of {\it
the} one sector which had not been tested before for the full current 
corresponding to an energy of $5$ TeV. Most likely, an electrical arc 
developed and it punctured the Helium enclosure.
Large amounts of Helium gas were released into the insulating vacuum of the 
cryostat and a large pressure wave traveled along the accelerator both ways.
The pressure was too large to handle for the pressure release systems which 
had been put in place. One of the quadrpole-dipole connection, before and 
after the  incident is shown in 
\begin{figure}
\begin{center}
\includegraphics*[scale=0.47]{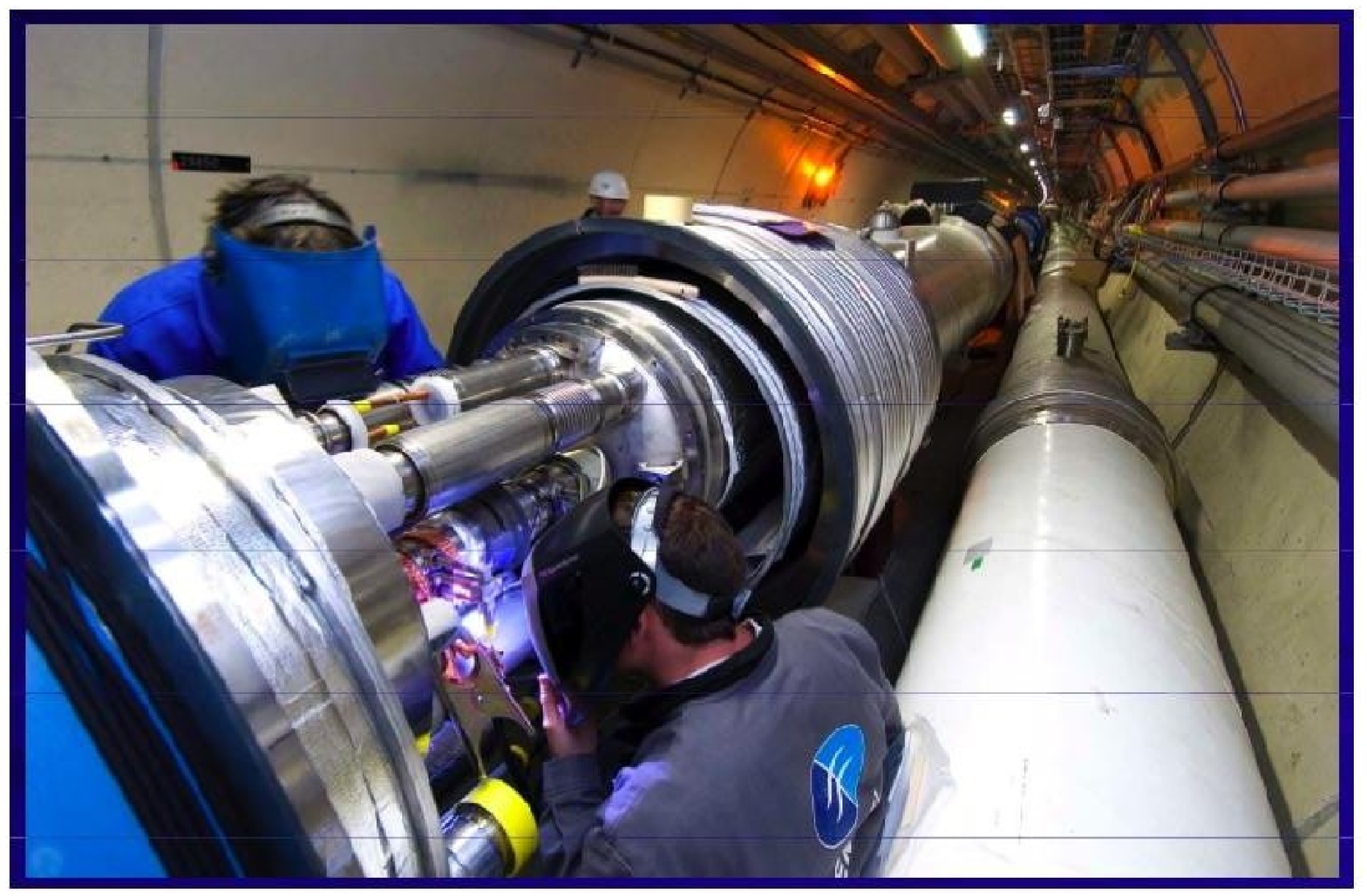}
\includegraphics*[scale=0.42]{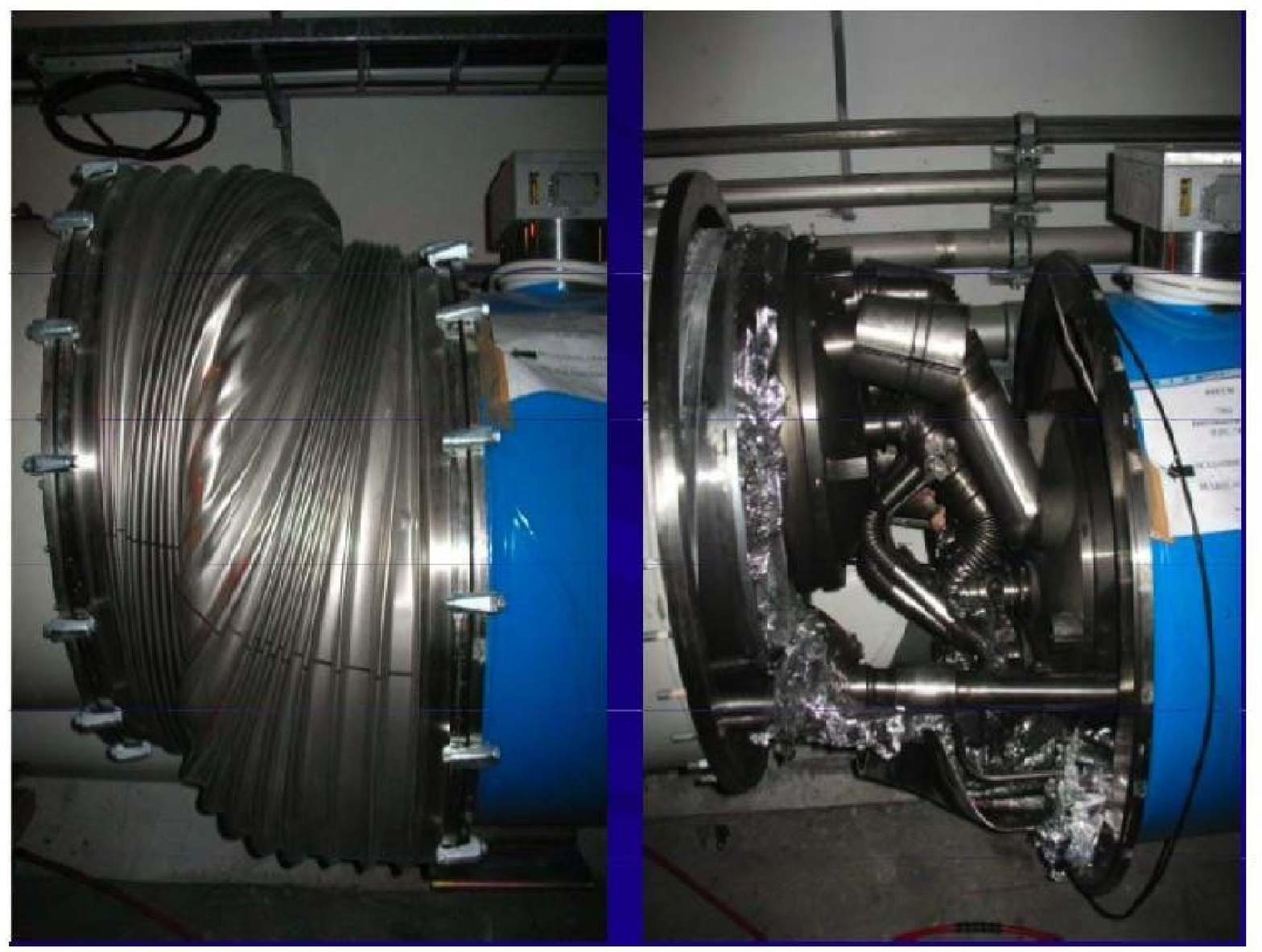}
\end{center}
\caption{The quadrupole-dipole connection: undamaged  and damaged in the 
left and right panel respectively.
}
\label{fig:cryodipole}
\end{figure}
Fig.~\ref{fig:cryodipole}. The severity of the accident becomes obvious when 
one recalls how heavy these magnets are. Thirty nine of the 1232 dipole magnets
and 16 of the quadrupole magnets had to be replaced, vacuum tubes had to be 
cleaned, new pressure release systems as well more diagnostic methods to avoid 
a similar accident again, had to be installed. All these repairs were 
completed by fall 2009. A large
amount of time is required to cool the machine to the low temperatures,
which has to be done gradually. Finally  by the end of the November 2009,
the machine was ready to go again and as said in the introduction the
collisions have taken place. However, it is noticed that the Superconducting
magnets need to be `trained' to carry higher currents and hence the
ramping of the energy will now be gradual. Initially now the machine will run 
only with beams of $3.5$ TeV each (exactly half the design energy and less than
the $5$ TeV which was planned for the 2009 run). It is now proposed that the 
machine will run at this energy till end of 2011 and then again there
will be  a major shutdown to ramp up the energy up to 6.5 to 7 TeV per beam,
(the 7 TeV looking a bit difficult to reach as per present studies), as well as 
an increase towards the ultimate design luminosity, of 
$10^{34} cm^{-2} sec^{-1}$. (This was the number that I had used in 
explaining how the LHC design energy was decided.) Now it is foreseen that 
the design energy and luminosity, will be only reached, 
after a period of about two years, as per current plan.

It should be mentioned that for this low energy run, a reduced luminosity of 
$10^{32} cm^{-2} sec^{-1}$ is foreseen. However, as I write these lines,
only a luminosity of $10^{30} cm^{-2} sec^{-1}$ has been achieved.
This is partially because these big machines are like a delicate musical 
instrument which needs to be finely tuned. For example, when larger
number of bunches are injected, to increase the luminosity, the 
bunch-bunch interactions can destabilise the beam. These interactions are 
understood, in principle, and included in the designs; still fine adjustments 
to the beam parameters are required to get rid of these effects while the 
machine is running. Right now  accelerator experts are working on it, 
alternating their work with giving the beam for actual physics run to the
experimental physicists. The bunches with the LHC design intensity (i.e.
$ 1.15\times 10^{11}$ p per bunch), have already been successfully injected. To 
increase the luminosity one would need to increase the number of bunches. After
beginning with one bunch, they had gone up to  13 bunches with $\sim 10^{10}$
ppb and 7 bunches for the $\sim 10^{11}$ ppb. The goal is to reach 2808 bunches
The possible time line of the expected physics results will in fact 
depend very much on the success of this tuning procedure. In the following 
I shall assume that in a few months at least a luminosity of few times 
$10^{31} cm^{-2} sec^{-1}$ will be achieved.

\subsection{Possible time line of Physics results}
As mentioned already, an evaluation of how well the mammoth detectors can
achieve their goals, requires precise theoretical predictions. A large number 
of very detailed analyses, with the combined participation of the theorists
and experimentalists alike were done to analyse the physics prospects of the
LHC~\cite{TDRS}. Most of the very detailed analyses used the nominal
LHC energy of $14$ TeV and the nominal luminosity of $10^{33}$ -- $10^{34}
cm^{-2} sec^{-1}$ in the first year. The value at the lower end means
that there will be $10$ events per year ($\sim 10^7$ sec = 150
running days in a year), per $fb$ cross-section. This is
termed as luminosity of $10$ fb$^{-1}$ per year.
\begin{figure}[hbt]
\begin{center}
\includegraphics*[height=11cm,width=9cm]{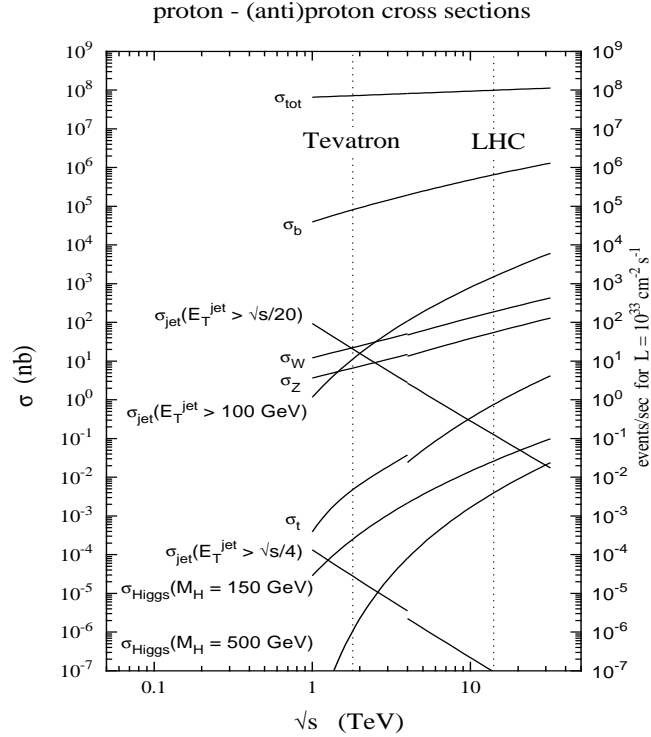}
\end{center}
\caption{The cross-sections for various processes expected at the LHC.}
\label{fig:lhccsec}
\end{figure}
The number on the right hand side indicates the number of events for
a particular process for the lower value of the nominal luminosity of 
$10 fb^{-1}$ or equivalently $10^{33}cm^{-2}sec^{-1}$.

One can see, from the figure that even at the high energy and luminosity, 
finding a Higgs with the lowest mass that is allowed for it by experimental 
constraints is not an easy job. For example, for a Higgs with $M_H c^2 = 150$ 
GeV, one expects only a handful of events where a Higgs boson is produced.
\footnote
{Further, when one multiplies it with the probability that it will give rise
to an event that can be distinguished from the background the number falls 
to may be an event or less a day.} 
One can see that the cross-sections for different processes
fall with varying factors as the collision energy is lowered. However, just a 
simple scaling by this factor is not enough to judge the change in sensitivity 
due to the present reduction in the energy and luminosity, as the background 
processes in general have a slower fall off with the reduction in energy 
than the signal process of interest. 

So in the short run, ie. in the next two years, with the machine delivering at
the most, a luminosity of 
$10^{32}cm^{-2} sec^{-1}$, one is not likely to see a signal for the SM 
Higgs boson, if it has mass in the range indicated by the current experimental 
constraints\footnote{It is interesting that already with the luminosity of 
$\sim 50$ 
nb$^{-1}$ that has been delivered the LHC could already have produced one event
of the type $p + p \rightarrow h + X \rightarrow b + \bar b +X$,  if its 
mass is around $120$ GeV. Unfortunately, the background is too huge in this 
channel and the ONLY promising channel for such a light Higgs, is when the $h$ 
decays into  $\gamma \gamma$  final state and such events will happen at a rate
1000 times smaller.} There is a possibility of detecting the Higgs, even for
this 'low' luminosity, if its mass is 
around  $160$--$165$ GeV, but the Tevatron data has ruled out existence of a 
SM Higgs in this mass range at $95\%$ c.l. However, due to limited statistics
and the theoretical uncertainty in the prediction of cross-section, this limit
has to be taken with some reservation~\cite{Baglio:2010um}.

Luckily, the production cross-section for the different varieties of new
particles that are expected in the different BSM scenarios, go down much more 
slowly with energy. For example, even with this lower luminosity and energy, 
the LHC has a chance to `discover' Supersymmetry if the masses of the strongly 
interacting super-particles are less than $800$ GeV, about twice the current 
limits from the Tevatron Collider. So clearly the possibilities for the BSM 
physics are quite tantalising even at the lower energy/lower luminosity LHC.
Thus the search for the Dark Matter in the Universe at the LHC can already 
happen in these two years. One good news is that the detectors are working
perfectly and are able to  remeasure the SM processes well, to calibrate
the SM backgrounds to the new physics searches using the experimental results
themselves.

The studies for the Quark Gluon Plasma formation (and hence recreating
the situations in the early Universe) should not be affected in a major way 
due to the energy reduction. Thus the heavy ion program and the $LHC_{b}$ 
program (to study the issue of CP violation and flavour physics at the LHC),
also should not be affected in a big way due to the reduced LHC conditions.

Whether the Higgs field is an elementary particle as in the SM or a`look-like'
which may not be elementary; whether there is only one or there are more of 
them : elementary or composite; one thing is for sure, the problem of 
reconciling a left handed world with massive matter particles, requires 
existence of a new phenomenon and the LHC has been designed to solve this 
enigma. However, we need to be patient, given the complexity of the problem, 
the rarity of the phenomena that we need to analyse to get the answers, along
with  the reduced energy and luminosity. Several years of data-taking and 
analysis will be needed to sort out the puzzle. 

It is heartening that Indian experimental groups have been contributing to 
building the detectors and doing experiments with the $pp$ as well as the 
heavy ion collisions. Further, the Indian theoretical
physics community is also involved in a big way in LHC physics, be it the 
SM, BSM physics in $pp$ collisions with the CMS/ATLAS detector, or the 
B-physics in $LHC_{b}$ or the physics of the QGP in the heavy ion mode.

After a shutdown that is planned at the end of 2011, the machine physicists
will work on getting to the higher energy. Once that energy is reached, LHC 
should be able to full fill its promise of hunting for the Higgs over the entire
mass range allowed for it theoretically, which is indeed the  `Raison de tr\'e'
of this machine. So in short this complicated exercise is sure to keep the
particle physics community busy for a decade or two and yield answers to some
of the basic questions about the very fabric of space and time and the
Universe we all live in, as the particle physicists travel in this 'terra
incognita'. In all probability it will throw up some unexpected results,
which in fact will point the way ahead in this journey towards truth. Exciting
times are ahead for sure !

\newpage
\begin{center}
{\large \bf  BOX1: Some relevant 
facts about the Standard Model (SM) of particle physics.
}
\end{center}
In the past 50-60 years particle physicists
have successfully arrived at a description of elementary constituents of 
matter the  matter particles quarks and leptons, all with  spin 1/2, 
\begin{table}[hbt]
\begin{center}
\begin{tabular}{|c|c|}
\hline
&\\
Quarks & Leptons \\
&\\
\hline
&\\
$\left( \begin{array}{c} u\\ d\\ \end{array} \right)$
\hspace{0.2cm}
$\left( \begin{array}{c} c\\ s\\ \end{array} \right)$
\hspace{0.2cm}
$\left( \begin{array}{c} t\\ b\\ \end{array} \right)$
&
$\left( \begin{array}{c} e^-\\ \nu_e\\ \end{array} \right)$
\hspace{0.2cm}
$\left( \begin{array}{c} \mu^-\\ \nu_\mu\\ \end{array} \right)$
\hspace{0.2cm}
$\left( \begin{array}{c} \tau^-\\ \nu_\tau\\ \end{array} \right)$
\\
$\times$ 3 colours & `colourless Leptons'\\
 & \\
\hspace{0.2cm}
${\color{red} \left( \begin{array}{c} u\\ d\\ \end{array} \right)}$
\hspace{0.2cm}
${\color{Olive} \left( \begin{array}{c} u\\ d\\ \end{array} \right)}$
\hspace{0.2cm}
${\color{Royal} \left( \begin{array}{c} u\\ d\\ \end{array} \right)}$
& `colourless Leptons' \\
{\it etc.}&\\
&\\
+anti-quarks & + anti-leptons \\ \hline
\end{tabular}
\caption{The fundamental constituents of matter.}
\label{tab1}
\end{center}
\end{table}
summarised in Table~\ref{tab1}, interacting with each other via the three basic interactions 
\begin{table}[bht]
\begin{center}
\begin{tabular}{|c|c|c|}
\hline
&&\\
Interaction&Description&Carrier Particle\\
&&\\
\hline
&&\\
Electromagnetic&Long-range
&Photon $\gamma$\\
&charged quarks and leptons.&\\
&&\\
Weak&Short-range
&$W/Z$ Bosons\\
& between quarks and leptons.&\\
&&\\
Strong&Short-range&Gluons $g$\\
&Only quarks&\\[3mm]
\hline
\end{tabular}
\caption{Basic forces in Nature and their carriers.}
\label{tab2}
\end{center}
\end{table}
shown in Table~\ref{tab2}~\footnote{The gravitational interaction is not 
included since we do not have a similar level of theoretical description of
this interaction in terms of a force carrier.}. The quarks possess the so 
called ''colour' charge and hence a given quark pair appears three times in the 
counting in three different colours. The quarks u,d (up,down) make the normal
matter (which is colourless) like protons/neutrons and these with the lightest 
charged lepton, the electron in fact make up atoms/molecules etc.  All the 
remaining fundamental particles: the quarks strange(s), charm (c), beauty (b) 
and top(t), the charged leptons $\mu,\tau$ and the neutral leptons: 
neutrinos ($\nu$'s), are produced either in decays of nuclei or unstable 
particles and/or in high energy processes.  The lighter quarks (u,d,s) manifest
themselves only as bound states like protons, pions and kaons. The heavier 
quarks and charged leptons are all short lived, with life times of the order of
$10^{-6}$ sec. or lower. The neutrinos have only weak interactions, whereas the
colourless charged leptons have weak and electromagnetic interactions and the 
coloured quarks feel all the three interactions. The properties of all the 
particles, the constituent matter particles and the force carriers, have been 
measured to a high degree of accuracy and the periodic table for particle 
physics is {\it almost} complete.  Hence, admittedly it is time to see if there
is an underlying theory which explains the patterns in these properties such as
their  masses and electromagnetic charge etc. that we observe. To answer such 
questions one has to go beyond the SM.

Tables~\ref{tab1} and~\ref{tab2} do not contain one  particle in the SM, viz. 
the Higgs boson, which is neither a fundamental constituent of matter nor a 
force carrier. It was introduced to understand a particular mystery of  weak 
interactions. These seem to treat the left handed matter particles, quark and 
leptons whose direction of the spin is opposite to the direction of motion, 
differently, from those which are right handed {\it i.e.} those for which
these two directions are parallel to each other. For a particle with a nonzero 
rest mass, a left handed state can be seen as a right handed state by simply 
going to a frame which is moving faster than the particle. Thus the weak 
interactions then will depend on the frame of reference. This would be in 
conflict with Einstein's theory of relativity. A theory with a Higgs boson 
does not suffer from this problem. This boson is named after one of the 
scientists who originally proposed this, more than 45 years ago. It is thought 
to have no electric charge, and no spin. As for its mass, unsuccessful 
searches at the LEP collider and the Tevatron collider, along with precise 
measurements of the weak interactions put $M_H c^2$ to be in the range of 
114 to $\sim$150 GeV, the result being strictly true within the SM. 

\newpage

\begin{center}
{\bf BOX2: A very brief summary of the theoretical proposals for BSM physics. 
}
\end{center}
A number of different ideas for BSM physics have been put forward through
the decades. They can be roughly classified into three classes:
\noindent 
1)The first class of models tries to keep the Higgs mass small by introduction 
of an additional symmetry. One of the most elegant ways to do this is via
Supersymmetry. This is a symmetry between fermions(spin 1/2 particles) and 
bosons (integral spin particles). This implies that there exist supersymmetric
partners of all the SM particles. In this case, there exist host of new 
particles which we should see at the colliders, particularly at the 
LHC~\cite{mybook} and it also has a DM candidate particle, the neutralino. 
The mass and interactions expected for the neutralino in SUSY models, falls 
in the range required to explain its abundance in the Universe today.   
Another class of models, called Little Higgs Models, in fact tries to use 
the lessons learned from the SIB in the case of the SM, to keep the Higgs light.
In this case also, there exist many additional fermions, gauge bosons in 
the theory at the TeV scale, their interaction patterns being different than in
the case of SUSY.

\noindent 2)The second class of models obviate the high energy scales which 
cause the theoretical predictions of corrections to Higgs mass to become large. These models are much more radical 
in that, in general they postulate behaviour of space and time which is 
completely different than what we understand and involve one or more extra
dimensions of space, which are compactified. These extra dimensions may even
have warped geometry.  In this case, the Higgs remains light, it may or 
may not be a fundamental particle.
Gravity is free to propagate in the extra  dimensions. 
Gravity in principle is as strong as the Electroweak interaction, but appears 
weak in our world.  TeV-scale experiments probe the
`strong’ gravity sector. Thus there is again new physics at a TeV scale.
Some of these ideas also make conceptual contact with 
quantum theory of gravitation and sometimes even have statements to make about 
early universe cosmology.  It is an exciting possibility, where the TeV energy
colliders can probe structure of space time. Even more interesting, string 
theory has now begun to make some statements about such models. 
In this case,
the Higgs remains light, it may or may not be a fundamental particle.

\noindent 3)Then there are extremely daring ideas which try to do without any
Higgs like particle; fundamental or composite.

It is  fair to say that in general all the models in class 2 and 3 above have 
more trouble satisfying the constraints coming from the precision measurements
than supersymmetry. On the other hand, not finding supersymmetric particles so 
far, has created another set of `theoretical' problems for supersymmetric 
theories which I will not go into here.

\end{document}